\documentclass[aps,preprint]{revtex4}%
\usepackage{amsfonts}
\usepackage{amsmath}
\usepackage{amssymb}
\usepackage{graphicx}%
\begin{document}
\title[Planck's Constant $h$]{Is Planck's Constant $h$ a "Quantum" Constant?}
\author{Timothy H. Boyer}
\affiliation{Department of Physics, City College of the City University of New York, New
York, New York 10031}

\begin{abstract}
One should not confuse a physical constant with a theory which incorporates
the constant. \ Planck's constant $h$ can appear in classical or quantum theories.

\end{abstract}
\maketitle

Physics students and sometimes physics teachers have a mistaken idea about the
role of Planck's constant in physics. \ Contrary to the impression of many,
the presence of Planck's constant $h$ in the formula for some physically
observed phenomenon does not necessarily indicate whether the formula was
derived from a classical or from a quantum theory. \ It seems important to
make a distinction between \textit{physical constants} and \textit{physical
theories}.

\textit{Physical constants} are numbers obtained from experimental
measurements. \ Examples are $g$=9.8m/sec$^{2}$ for the acceleration due to
gravity near the surface of the earth, $c$=3.0x10$^{8}$m/sec for the speed of
light in vacuum, $e$=1.6x10$^{-19}$C for the smallest elementary charge,
$a_{S}$=7.6x10$^{-16}$J/(m$^{3}$K$^{4})$ for Stefan's radiation constant,
$h$=6.6x10$^{-34}$J\thinspace sec for Planck's constant, and $v_{air}%
$=344m/sec for the speed of sound in air at normal pressure and density.
\ \ Some of these constants, such as $c$, $e$, $a_{S},$ and $h,$ are regarded
as universal, whereas others, such as $g$ and $v_{air}$, depend upon very
specific experimental circumstances.

\textit{Physical Theories} are sets of rules for describing physical
phenomena; these theories may or may not incorporate various physical
constants. \ Thus, for example, a Newtonian theory of particle motion near the
surface of the earth which does incorporate gravity would treat $mg$ as a
downward force on a mass $m$ and would give vertical free-particle motion as
\begin{equation}
y=y_{0}+v_{0}t-(1/2)gt^{2}%
\end{equation}
(measuring displacement vertically upwards). \ On the other hand, a classical
mechanical theory which did \textit{not} incorporate gravity might give the
motion as%
\begin{equation}
y=y_{0}+v_{0}t
\end{equation}
omitting the physical constant $g$, and so giving a less accurate description
of the phenomenon. \ Equation (2) becomes accurate only at high initial
velocities and short times when the effects associated with the physical
constant $g$ can be ignored. \ Finally, the physical situation can also be
described by general relativity, in which case the physical constant $g$ would
again enter the theory; however, in this case, the constant $g$ would be
related to the curvature of spacetime. \ Within the appropriate approximation,
the general relativistic analysis would lead back to Eq. (1). \ In this
example, both Newtonian gravity and general relativity incorporate the
physical constant $g,$ but incorporate it in very different fashions. \ The
mere presence of the physical constant $g$ in Eq. (1) does not indicate which
theory was used as the starting point in the physical analysis.

An analogous situation holds for the physical constant $h,$ Planck's constant.
\ \textit{Traditional} classical mechanics and \textit{traditional} classical
electrodynamics do not include the physical constant $h,$ just as the
classical description above which omitted gravity does not include $g.$
\ Accordingly, these theories give accurate descriptions of phenomena only
when the constants $h$ or $g$ play no significant role. \ However, there are
several theories which do include $h$, and the final formula which describes
some phenomenon does not indicate which theory was used as the starting point
for the analysis. \ Among the physical theories which include Planck's
constant $h$ are nonrelativistic quantum mechanics, quantum electrodynamics,
and classical electrodynamics with classical electromagnetic zero-point
radiation (stochastic electrodynamics).\cite{M} \ Of course, quantum mechanics
and quantum electrodynamics are widely-known successful theories which are
taught in universities. \ However, classical electrodynamics with classical
electromagnetic zero-point radiation is a mathematically-consistent classical
theory which incorporates Planck's constant $h$ as the scale constant of the
Lorentz-invariant spectrum of random classical radiation which is chosen as
the homogeneous boundary condition on Maxwell's equations. \ The zero-point
radiation in this theory is described in the same terms as the thermal
radiation of nineteenth century physics. \ There are no discrete (quantum)
aspects of energy or momentum in this classical theory. \ In
\textit{traditional} classical electrodynamics, the homogeneous solution of
Maxwell's equations is taken to vanish so that there is no zero-point
radiation, and therefore Planck's constant $h$ simply does not enter the
theory at all.\cite{Lorentz} \ Classical electrodynamics with classical
electromagnetic zero-point radiation has been shown to give the same results
as nonrelativistic quantum mechanics and quantum electrodynamics for free
electromagnetic fields and for linear mechanical systems weakly coupled to
radiation.\cite{B} \ The classical theory can account for Casimir forces
between macroscopic objects, van der Waals forces between molecules,
oscillator specific heats, specific heats of solids, diamagnetism, and the
thermal effects of acceleration through the vacuum, giving results identical
to those from quantum theories.\cite{PC} \ Also, simulation work indicates
that this classical theory can account for the Schroedinger ground state of
hydrogen.\cite{C} \ The limits of the theory are still being
explored.\cite{B2} \ Most importantly, the appearance of Planck's constant $h$
in the final formula for these phenomena does not indicate whether the
starting theory was a classical or a quantum theory.

\end{document}